\documentclass[conference,10pt,twoside,twocolumn]{IEEEtran}
\IEEEoverridecommandlockouts
\usepackage{cite}
\usepackage[T1]{fontenc}
\usepackage{graphicx}
\usepackage{amssymb}
\usepackage{amsmath}
\usepackage{amsthm}
\usepackage{subfigure}
\usepackage{booktabs} %\toprule \midrule \bottomrule
\usepackage{multirow}
\usepackage{microtype}
\usepackage{balance}
\usepackage{xcolor}
\usepackage{fancyref}
\usepackage{courier}
\usepackage{dblfloatfix}
\usepackage{listings}

\usepackage[hidelinks]{hyperref}
\usepackage{amssymb}
\usepackage{amsfonts}
\usepackage{mathrsfs}
\usepackage{xspace}
\usepackage{bm}
\usepackage{upgreek}

\newcommand{\safemath}[2]{\newcommand{#1}{\ensuremath{#2}\xspace}}

\safemath{\bma}{\mathbf{a}}
\safemath{\bmb}{\mathbf{b}}
\safemath{\bmc}{\mathbf{c}}
\safemath{\bmd}{\mathbf{d}}
\safemath{\bme}{\mathbf{e}}
\safemath{\bmf}{\mathbf{f}}
\safemath{\bmg}{\mathbf{g}}
\safemath{\bmh}{\mathbf{h}}
\safemath{\bmi}{\mathbf{i}}
\safemath{\bmj}{\mathbf{j}}
\safemath{\bmk}{\mathbf{k}}
\safemath{\bml}{\mathbf{l}}
\safemath{\bmm}{\mathbf{m}}
\safemath{\bmn}{\mathbf{n}}
\safemath{\bmo}{\mathbf{o}}
\safemath{\bmp}{\mathbf{p}}
\safemath{\bmq}{\mathbf{q}}
\safemath{\bmr}{\mathbf{r}}
\safemath{\bms}{\mathbf{s}}
\safemath{\bmt}{\mathbf{t}}
\safemath{\bmu}{\mathbf{u}}
\safemath{\bmv}{\mathbf{v}}
\safemath{\bmw}{\mathbf{w}}
\safemath{\bmx}{\mathbf{x}}
\safemath{\bmy}{\mathbf{y}}
\safemath{\bmz}{\mathbf{z}}
\safemath{\bmzero}{\mathbf{0}}
\safemath{\bmone}{\mathbf{1}}
\safemath{\Bell}{\ensuremath{\boldsymbol\ell}}

\bmdefine{\biad}{a}
\bmdefine{\bibd}{b}
\bmdefine{\bicd}{c}
\bmdefine{\bidd}{d}
\bmdefine{\bied}{e}
\bmdefine{\bifd}{f}
\bmdefine{\bigd}{g}
\bmdefine{\bihd}{h}
\bmdefine{\biid}{i}
\bmdefine{\bijd}{j}
\bmdefine{\bikd}{k}
\bmdefine{\bild}{l}
\bmdefine{\bimd}{m}
\bmdefine{\bind}{n}
\bmdefine{\biod}{o}
\bmdefine{\bipd}{p}
\bmdefine{\biqd}{q}
\bmdefine{\bird}{r}
\bmdefine{\bisd}{s}
\bmdefine{\bitd}{t}
\bmdefine{\biud}{u}
\bmdefine{\bivd}{v}
\bmdefine{\biwd}{w}
\bmdefine{\bixd}{x}
\bmdefine{\biyd}{y}
\bmdefine{\bizd}{z}

\bmdefine{\bixid}{\xi}
\bmdefine{\bilambdad}{\lambda}
\bmdefine{\bimud}{\mu}
\bmdefine{\bithetad}{\theta}
\bmdefine{\biphid}{\phi}
\bmdefine{\bideltad}{\delta}

\safemath{\bmia}{\biad}
\safemath{\bmib}{\bibd}
\safemath{\bmic}{\bicd}
\safemath{\bmid}{\bidd}
\safemath{\bmie}{\bied}
\safemath{\bmif}{\bifd}
\safemath{\bmig}{\bigd}
\safemath{\bmih}{\bihd}
\safemath{\bmii}{\biid}
\safemath{\bmij}{\bijd}
\safemath{\bmik}{\bikd}
\safemath{\bmil}{\bild}
\safemath{\bmim}{\bimd}
\safemath{\bmin}{\bind}
\safemath{\bmio}{\biod}
\safemath{\bmip}{\bipd}
\safemath{\bmiq}{\biqd}
\safemath{\bmir}{\bird}
\safemath{\bmis}{\bisd}
\safemath{\bmit}{\bitd}
\safemath{\bmiu}{\biud}
\safemath{\bmiv}{\bivd}
\safemath{\bmiw}{\biwd}
\safemath{\bmix}{\bixd}
\safemath{\bmiy}{\biyd}
\safemath{\bmiz}{\bizd}

\safemath{\bmxi}{\bixid}
\safemath{\bmlambda}{\bilambdad}
\safemath{\bmmu}{\bimud}
\safemath{\bmtheta}{\bithetad}
\safemath{\bmphi}{\biphid}
\safemath{\bmdelta}{\bideltad}

\safemath{\bA}{\mathbf{A}}
\safemath{\bB}{\mathbf{B}}
\safemath{\bC}{\mathbf{C}}
\safemath{\bD}{\mathbf{D}}
\safemath{\bE}{\mathbf{E}}
\safemath{\bF}{\mathbf{F}}
\safemath{\bG}{\mathbf{G}}
\safemath{\bH}{\mathbf{H}}
\safemath{\bI}{\mathbf{I}}
\safemath{\bJ}{\mathbf{J}}
\safemath{\bK}{\mathbf{K}}
\safemath{\bL}{\mathbf{L}}
\safemath{\bM}{\mathbf{M}}
\safemath{\bN}{\mathbf{N}}
\safemath{\bO}{\mathbf{O}}
\safemath{\bP}{\mathbf{P}}
\safemath{\bQ}{\mathbf{Q}}
\safemath{\bR}{\mathbf{R}}
\safemath{\bS}{\mathbf{S}}
\safemath{\bT}{\mathbf{T}}
\safemath{\bU}{\mathbf{U}}
\safemath{\bV}{\mathbf{V}}
\safemath{\bW}{\mathbf{W}}
\safemath{\bX}{\mathbf{X}}
\safemath{\bY}{\mathbf{Y}}
\safemath{\bZ}{\mathbf{Z}}

\safemath{\bZero}{\mathbf{0}}
\safemath{\bOne}{\mathbf{1}}
\safemath{\bDelta}{\mathbf{\Delta}}
\safemath{\bLambda}{\mathbf{\UpLambda}}
\safemath{\bPhi}{\mathbf{\Upphi}}
\safemath{\bSigma}{\mathbf{\Upsigma}}
\safemath{\bOmega}{\mathbf{\Upomega}}
\safemath{\bTheta}{\mathbf{\Uptheta}}

\bmdefine{\biAd}{A}
\bmdefine{\biBd}{B}
\bmdefine{\biCd}{C}
\bmdefine{\biDd}{D}
\bmdefine{\biEd}{E}
\bmdefine{\biFd}{F}
\bmdefine{\biGd}{G}
\bmdefine{\biHd}{H}
\bmdefine{\biId}{I}
\bmdefine{\biJd}{J}
\bmdefine{\biKd}{K}
\bmdefine{\biLd}{L}
\bmdefine{\biMd}{M}
\bmdefine{\biOd}{N}
\bmdefine{\biPd}{O}
\bmdefine{\biQd}{P}
\bmdefine{\biRd}{R}
\bmdefine{\biSd}{S}
\bmdefine{\biTd}{T}
\bmdefine{\biUd}{U}
\bmdefine{\biVd}{V}
\bmdefine{\biWd}{W}
\bmdefine{\biXd}{X}
\bmdefine{\biYd}{Y}
\bmdefine{\biZd}{Z}

\bmdefine{\biDelta}{\Delta}
\bmdefine{\biLambda}{\Lambda}
\bmdefine{\biPhi}{\Phi}
\bmdefine{\biSigma}{\Sigma}
\bmdefine{\biOmega}{\Omega}
\bmdefine{\biTheta}{\Theta}

\safemath{\bimA}{\biAd}
\safemath{\bimB}{\biBd}
\safemath{\bimC}{\biCd}
\safemath{\bimD}{\biDd}
\safemath{\bimE}{\biEd}
\safemath{\bimF}{\biFd}
\safemath{\bimG}{\biGd}
\safemath{\bimH}{\biHd}
\safemath{\bimI}{\biId}
\safemath{\bimJ}{\biJd}
\safemath{\bimK}{\biKd}
\safemath{\bimL}{\biLd}
\safemath{\bimM}{\biMd}
\safemath{\bimN}{\biNd}
\safemath{\bimO}{\biOd}
\safemath{\bimP}{\biPd}
\safemath{\bimQ}{\biQd}
\safemath{\bimR}{\biRd}
\safemath{\bimS}{\biSd}
\safemath{\bimT}{\biTd}
\safemath{\bimU}{\biUd}
\safemath{\bimV}{\biVd}
\safemath{\bimW}{\biWd}
\safemath{\bimX}{\biXd}
\safemath{\bimY}{\biYd}
\safemath{\bimZ}{\biZd}

\safemath{\bimDelta}{\biDelta}
\safemath{\bimLambda}{\biLambda}
\safemath{\bimPhi}{\biPhi}
\safemath{\bimSigma}{\biSigma}
\safemath{\bimOmega}{\biOmega}
\safemath{\bimTheta}{\biTheta}

\safemath{\setA}{\mathcal{A}}
\safemath{\setB}{\mathcal{B}}
\safemath{\setC}{\mathcal{C}}
\safemath{\setD}{\mathcal{D}}
\safemath{\setE}{\mathcal{E}}
\safemath{\setF}{\mathcal{F}}
\safemath{\setG}{\mathcal{G}}
\safemath{\setH}{\mathcal{H}}
\safemath{\setI}{\mathcal{I}}
\safemath{\setJ}{\mathcal{J}}
\safemath{\setK}{\mathcal{K}}
\safemath{\setL}{\mathcal{L}}
\safemath{\setM}{\mathcal{M}}
\safemath{\setN}{\mathcal{N}}
\safemath{\setO}{\mathcal{O}}
\safemath{\setP}{\mathcal{P}}
\safemath{\setQ}{\mathcal{Q}}
\safemath{\setR}{\mathcal{R}}
\safemath{\setS}{\mathcal{S}}
\safemath{\setT}{\mathcal{T}}
\safemath{\setU}{\mathcal{U}}
\safemath{\setV}{\mathcal{V}}
\safemath{\setW}{\mathcal{W}}
\safemath{\setX}{\mathcal{X}}
\safemath{\setY}{\mathcal{Y}}
\safemath{\setZ}{\mathcal{Z}}
\safemath{\emptySet}{\varnothing}

\safemath{\colA}{\mathscr{A}}
\safemath{\colB}{\mathscr{B}}
\safemath{\colC}{\mathscr{C}}
\safemath{\colD}{\mathscr{D}}
\safemath{\colE}{\mathscr{E}}
\safemath{\colF}{\mathscr{F}}
\safemath{\colG}{\mathscr{G}}
\safemath{\colH}{\mathscr{H}}
\safemath{\colI}{\mathscr{I}}
\safemath{\colJ}{\mathscr{J}}
\safemath{\colK}{\mathscr{K}}
\safemath{\colL}{\mathscr{L}}
\safemath{\colM}{\mathscr{M}}
\safemath{\colN}{\mathscr{N}}
\safemath{\colO}{\mathscr{O}}
\safemath{\colP}{\mathscr{P}}
\safemath{\colQ}{\mathscr{Q}}
\safemath{\colR}{\mathscr{R}}
\safemath{\colS}{\mathscr{S}}
\safemath{\colT}{\mathscr{T}}
\safemath{\colU}{\mathscr{U}}
\safemath{\colV}{\mathscr{V}}
\safemath{\colW}{\mathscr{W}}
\safemath{\colX}{\mathscr{X}}
\safemath{\colY}{\mathscr{Y}}
\safemath{\colZ}{\mathscr{Z}}

\safemath{\opA}{\mathbb{A}}
\safemath{\opB}{\mathbb{B}}
\safemath{\opC}{\mathbb{C}}
\safemath{\opD}{\mathbb{D}}
\safemath{\opE}{\mathbb{E}}
\safemath{\opF}{\mathbb{F}}
\safemath{\opG}{\mathbb{G}}
\safemath{\opH}{\mathbb{H}}
\safemath{\opI}{\mathbb{I}}
\safemath{\opJ}{\mathbb{J}}
\safemath{\opK}{\mathbb{K}}
\safemath{\opL}{\mathbb{L}}
\safemath{\opM}{\mathbb{M}}
\safemath{\opN}{\mathbb{N}}
\safemath{\opO}{\mathbb{O}}
\safemath{\opP}{\mathbb{P}}
\safemath{\opQ}{\mathbb{Q}}
\safemath{\opR}{\mathbb{R}}
\safemath{\opS}{\mathbb{S}}
\safemath{\opT}{\mathbb{T}}
\safemath{\opU}{\mathbb{U}}
\safemath{\opV}{\mathbb{V}}
\safemath{\opW}{\mathbb{W}}
\safemath{\opX}{\mathbb{X}}
\safemath{\opY}{\mathbb{Y}}
\safemath{\opZ}{\mathbb{Z}}
\safemath{\opZero}{\mathbb{O}}
\safemath{\identityop}{\opI}

\safemath{\veca}{\bma}
\safemath{\vecb}{\bmb}
\safemath{\vecc}{\bmc}
\safemath{\vecd}{\bmd}
\safemath{\vece}{\bme}
\safemath{\vecf}{\bmf}
\safemath{\vecg}{\bmg}
\safemath{\vech}{\bmh}
\safemath{\veci}{\bmi}
\safemath{\vecj}{\bmj}
\safemath{\veck}{\bmk}
\safemath{\vecl}{\bml}
\safemath{\vecm}{\bmm}
\safemath{\vecn}{\bmn}
\safemath{\veco}{\bmo}
\safemath{\vecp}{\bmp}
\safemath{\vecq}{\bmq}
\safemath{\vecr}{\bmr}
\safemath{\vecs}{\bms}
\safemath{\vect}{\bmt}
\safemath{\vecu}{\bmu}
\safemath{\vecv}{\bmv}
\safemath{\vecw}{\bmw}
\safemath{\vecx}{\bmx}
\safemath{\vecy}{\bmy}
\safemath{\vecz}{\bmz}

\safemath{\veczero}{\bmzero}
\safemath{\vecone}{\bmone}
\safemath{\vecxi}{\bmxi}
\safemath{\veclambda}{\bmlambda}
\safemath{\vecmu}{\bmmu}
\safemath{\vectheta}{\bmtheta}
\safemath{\vecphi}{\bmphi}
\safemath{\vecdelta}{\bmdelta}

\safemath{\matA}{\bA}
\safemath{\matB}{\bB}
\safemath{\matC}{\bC}
\safemath{\matD}{\bD}
\safemath{\matE}{\bE}
\safemath{\matF}{\bF}
\safemath{\matG}{\bG}
\safemath{\matH}{\bH}
\safemath{\matI}{\bI}
\safemath{\matJ}{\bJ}
\safemath{\matK}{\bK}
\safemath{\matL}{\bL}
\safemath{\matM}{\bM}
\safemath{\matN}{\bN}
\safemath{\matO}{\bO}
\safemath{\matP}{\bP}
\safemath{\matQ}{\bQ}
\safemath{\matR}{\bR}
\safemath{\matS}{\bS}
\safemath{\matT}{\bT}
\safemath{\matU}{\bU}
\safemath{\matV}{\bV}
\safemath{\matW}{\bW}
\safemath{\matX}{\bX}
\safemath{\matY}{\bY}
\safemath{\matZ}{\bZ}
\safemath{\matzero}{\bmzero}

\safemath{\matDelta}{\bDelta}
\safemath{\matLambda}{\bLambda}
\safemath{\matPhi}{\bPhi}
\safemath{\matSigma}{\bSigma}
\safemath{\matOmega}{\bOmega}
\safemath{\matTheta}{\bTheta}

\safemath{\matidentity}{\matI}
\safemath{\matone}{\matO}

\safemath{\rnda}{A}
\safemath{\rndb}{B}
\safemath{\rndc}{C}
\safemath{\rndd}{D}
\safemath{\rnde}{E}
\safemath{\rndf}{F}
\safemath{\rndg}{G}
\safemath{\rndh}{H}
\safemath{\rndi}{I}
\safemath{\rndj}{J}
\safemath{\rndk}{K}
\safemath{\rndl}{L}
\safemath{\rndm}{M}
\safemath{\rndn}{N}
\safemath{\rndo}{O}
\safemath{\rndp}{P}
\safemath{\rndq}{Q}
\safemath{\rndr}{R}
\safemath{\rnds}{S}
\safemath{\rndt}{T}
\safemath{\rndu}{U}
\safemath{\rndv}{V}
\safemath{\rndw}{W}
\safemath{\rndx}{X}
\safemath{\rndy}{Y}
\safemath{\rndz}{Z}

\safemath{\rveca}{\bimA}
\safemath{\rvecb}{\bimB}
\safemath{\rvecc}{\bimC}
\safemath{\rvecd}{\bimD}
\safemath{\rvece}{\bimE}
\safemath{\rvecf}{\bimF}
\safemath{\rvecg}{\bimG}
\safemath{\rvech}{\bimH}
\safemath{\rveci}{\bimI}
\safemath{\rvecj}{\bimJ}
\safemath{\rveck}{\bimK}
\safemath{\rvecl}{\bimL}
\safemath{\rvecm}{\bimM}
\safemath{\rvecn}{\bimN}
\safemath{\rveco}{\bomO}
\safemath{\rvecp}{\bimP}
\safemath{\rvecq}{\bimQ}
\safemath{\rvecr}{\bimR}
\safemath{\rvecs}{\bimS}
\safemath{\rvect}{\bimT}
\safemath{\rvecu}{\bimU}
\safemath{\rvecv}{\bimV}
\safemath{\rvecw}{\bimW}
\safemath{\rvecx}{\bimX}
\safemath{\rvecy}{\bimY}
\safemath{\rvecz}{\bimZ}

\safemath{\rvecxi}{\bmxi}
\safemath{\rveclambda}{\bmlambda}
\safemath{\rvecmu}{\bmmu}
\safemath{\rvectheta}{\bmtheta}
\safemath{\rvecphi}{\bmphi}

\safemath{\rmatA}{\bimA}
\safemath{\rmatB}{\bimB}
\safemath{\rmatC}{\bimC}
\safemath{\rmatD}{\bimD}
\safemath{\rmatE}{\bimE}
\safemath{\rmatF}{\bimF}
\safemath{\rmatG}{\bimG}
\safemath{\rmatH}{\bimH}
\safemath{\rmatI}{\bimI}
\safemath{\rmatJ}{\bimJ}
\safemath{\rmatK}{\bimK}
\safemath{\rmatL}{\bimL}
\safemath{\rmatM}{\bimM}
\safemath{\rmatN}{\bimN}
\safemath{\rmatO}{\bimO}
\safemath{\rmatP}{\bimP}
\safemath{\rmatQ}{\bimQ}
\safemath{\rmatR}{\bimR}
\safemath{\rmatS}{\bimS}
\safemath{\rmatT}{\bimT}
\safemath{\rmatU}{\bimU}
\safemath{\rmatV}{\bimV}
\safemath{\rmatW}{\bimW}
\safemath{\rmatX}{\bimX}
\safemath{\rmatY}{\bimY}
\safemath{\rmatZ}{\bimZ}

\safemath{\rmatDelta}{\bimDelta}
\safemath{\rmatLambda}{\bimLambda}
\safemath{\rmatPhi}{\bimPhi}
\safemath{\rmatSigma}{\bimSigma}
\safemath{\rmatOmega}{\bimOmega}
\safemath{\rmatTheta}{\bimTheta}

\usepackage{amssymb}
\usepackage{amsfonts}
\usepackage{mathrsfs}
\usepackage{xspace}
\usepackage{bm}
\usepackage{fancyref}
\usepackage{textcomp}

\usepackage{multirow}
\usepackage{stmaryrd}

\newenvironment{textbmatrix}{	\setlength{\arraycolsep}{2.5pt}%
								\left[\begin{matrix}}{\end{matrix}\right]%
								\raisebox{0.08ex}{\vphantom{M}}}

\def\be{\begin{equation}}
\def\ee{\end{equation}}
\def\een{\nonumber \end{equation}}
\def\mat{\begin{bmatrix}}
\def\emat{\end{bmatrix}}
\def\btm{\begin{textbmatrix}}
\def\etm{\end{textbmatrix}}

\def\ba#1\ea{\begin{align}#1\end{align}}
\def\bas#1\eas{\begin{align*}#1\end{align*}}
\def\bs#1\es{\begin{split}#1\end{split}}
\def\bg#1\eg{\begin{gather}#1\end{gather}}
\def\bml#1\eml{\begin{multline}#1\end{multline}}
\def\bi#1\ei{\begin{itemize}#1\end{itemize}}

\newcommand{\pinv}[1]{\ensuremath{#1^{\dagger}}} 	% Moore-Penrose pseudo-inverse
\safemath{\dirac}{\delta}					% Dirac delta
\safemath{\krond}{\dirac}					% Kronecker delta
\safemath{\upto}{\uparrow}
\safemath{\downto}{\downarrow}
\safemath{\iu}{j}							% imaginary unit
\safemath{\ev}{\lambda}						% eigenvalue
\safemath{\hilseqspace}{l^{2}}				% Hilbert sequence space
\newcommand{\banachfunspace}[1]{\setL^{#1}}	% Banach function space
\safemath{\hilfunspace}{\banachfunspace{2}}	% Hilbert function space
\safemath{\SNR}{\textit{SNR}} 				% signal to noise ratio
\safemath{\PAR}{\textit{PAR}} 				% signal to noise ratio
\safemath{\No}{N_0}							% noise spectral density
\safemath{\Es}{E_s}							% energy per symbol
\safemath{\Eb}{E_b}							% energy per bit
\safemath{\EbNo}{\frac{\Eb}{\No}}
\safemath{\EsNo}{\frac{\Es}{\No}}

\DeclareMathOperator{\CHop}{\ensuremath{\opH}} % channel operator
\safemath{\tvir}{\rndh_{\CHop}}				% time-varying impulse response
\safemath{\tvtf}{\rndl_{\CHop}}				% 	-''- transfer function
\safemath{\spf}{\rnds_{\CHop}}				% spreading function
\safemath{\bff}{H_{\CHop}}					% bi-freuqency function

\safemath{\ircf}{r_{h}}						% impulse response correlation fn.
\safemath{\tftvcf}{r_{s}}					% scattering function
\safemath{\tfcf}{r_{l}}						% time-frequency correlation fn.
\safemath{\bfcf}{r_{H}}						% bi-frequency correlation fn.

\safemath{\tcorr}{c_h}						% time-correlation function
\safemath{\scf}{c_{s}}						% spreading function
\safemath{\tfcorr}{c_{l}}					% transfer-function correlation
\safemath{\fcorr}{c_{H}}						% frequency-correlation function

\safemath{\mi}{I}							% mutual information
\safemath{\capacity}{C}						% capacity

\safemath{\normal}{\mathcal{N}}			% normal distribution
\safemath{\jpg}{\mathcal{CN}}			% jointly proper Gaussian
\safemath{\mchain}{\leftrightarrow}		% Markov chain
\safemath{\dB}{\,\mathrm{dB}}
\safemath{\dBm}{\,\mathrm{dBm}}
\safemath{\Hz}{\,\mathrm{Hz}}
\safemath{\kHz}{\,\mathrm{kHz}}
\safemath{\MHz}{\,\mathrm{MHz}}
\safemath{\GHz}{\,\mathrm{GHz}}
\safemath{\s}{\,\mathrm{s}}
\safemath{\ms}{\,\mathrm{ms}}
\safemath{\mus}{\,\mathrm{\text{\textmu}s}}
\safemath{\ns}{\,\mathrm{ns}}
\safemath{\ps}{\,\mathrm{ps}}
\safemath{\meter}{\,\mathrm{m}}
\safemath{\mm}{\,\mathrm{mm}}
\safemath{\cm}{\,\mathrm{cm}}
\safemath{\m}{\,\mathrm{m}}
\safemath{\W}{\,\mathrm{W}}
\safemath{\mW}{\, \mathrm{mW}}
\safemath{\J}{\,\mathrm{J}}
\safemath{\K}{\,\mathrm{K}}
\safemath{\bit}{\,\mathrm{bit}}
\safemath{\nat}{\,\mathrm{nat}}

\safemath{\define}{\triangleq}			% definition

\safemath{\equivalent}{\sim}
\safemath{\distas}{\sim}					% distributed according to
\safemath{\sdiff}{\Delta}				% symmetric set difference

\safemath{\reals}{\mathbb{R}}
\safemath{\positivereals}{\reals_{+}}
\safemath{\integers}{\mathbb{Z}}
\safemath{\posint}{\integers_{+}}
\safemath{\naturals}{\mathbb{N}}
\safemath{\posnaturals}{\naturals_{+}}
\safemath{\complexset}{\mathbb{C}}
\safemath{\rationals}{\mathbb{Q}}

\newcommand*{\fancyrefapplabelprefix}{app}		% Appendix
\newcommand*{\fancyrefthmlabelprefix}{thm}		% Theorem
\newcommand*{\fancyreflemlabelprefix}{lem}		% Lemma
\newcommand*{\fancyrefcorlabelprefix}{cor}		% Corollary
\newcommand*{\fancyrefdeflabelprefix}{def}		% Definition
\newcommand*{\fancyrefproplabelprefix}{prop}		% Proposition
\newcommand*{\fancyrefexmpllabelprefix}{exmpl}
\newcommand*{\fancyrefalglabelprefix}{alg}		% Algorithm
\newcommand*{\fancyreftbllabelprefix}{tbl}		% Algorithm

\frefformat{vario}{\fancyrefseclabelprefix}{Sec.~#1}
\frefformat{vario}{\fancyrefthmlabelprefix}{Thm.~#1}
\frefformat{vario}{\fancyreftbllabelprefix}{Tbl.~#1}
\frefformat{vario}{\fancyreflemlabelprefix}{Lem.~#1}
\frefformat{vario}{\fancyrefcorlabelprefix}{Corr.~#1}
\frefformat{vario}{\fancyrefdeflabelprefix}{Def.~#1}
\frefformat{vario}{\fancyreffiglabelprefix}{Fig.~#1}
\frefformat{vario}{\fancyrefapplabelprefix}{App.~#1}
\frefformat{vario}{\fancyrefeqlabelprefix}{(#1)}
\frefformat{vario}{\fancyrefproplabelprefix}{Prop.~#1}
\frefformat{vario}{\fancyrefexmpllabelprefix}{Ex.~#1}
\frefformat{vario}{\fancyrefalglabelprefix}{Alg.~#1}

\safemath{\dictab}{[\,\dicta\,\,\dictb\,]}

\safemath{\ysig}{\bmy}
\safemath{\ysighat}{\hat{\ysig}}
\safemath{\ysigdim}{M}
\safemath{\xsig}{\bmx}
\safemath{\xsigdim}{N}
\safemath{\nx}{n_x}
\safemath{\zsig}{\bmz}
\safemath{\zsigdim}{\ysigdim}
\safemath{\rsig}{\bmr}
\safemath{\Adict}{\bA}
\safemath{\Adicttilde}{\widetilde{\Adict}}
\safemath{\Adictdim}{\outputdim\times\xsigdim}
\safemath{\avec}{\bma}
\safemath{\avectilde}{\tilde{\avec}}
\safemath{\Bdict}{\bB}
\safemath{\Bdicttilde}{\widetilde{\Bdict}}
\safemath{\Cdict}{\bC}
\safemath{\cvec}{\bmc}
\safemath{\Ddict}{\bD}
\safemath{\Ddictdim}{\ysigdim\times\xsigdim}
\safemath{\dvec}{\bmd}
\safemath{\Ddicttilde}{\widetilde{\bD}}
\safemath{\Bonb}{\bB}
\safemath{\bvec}{\bmb}
\safemath{\Bonbdim}{\ysigdim\times\ysigdim}
\safemath{\noise}{\bmn}
\safemath{\noisedim}{\ysigim}
\safemath{\err}{\bme}
\safemath{\errdim}{\ysigdim}
\safemath{\errset}{\setE}
\safemath{\nerr}{n_e}
\safemath{\delop}{\bP_\errset}
\safemath{\delopc}{\bP_{{\errset}^c}}

\safemath{\cplxi}{\imath}
\safemath{\cplxj}{\jmath}
\safemath{\dict}{\matD}
\safemath{\inputdim}{N}		% number of columns of dictionary D
\safemath{\outputdim}{M}		%number of rows of dictionary D
\safemath{\sparsity}{S}	%sparsity
\safemath{\inputdimA}{{N_a}}	%total number of elements in dictionary A
\safemath{\inputdimB}{{N_b}}	%total number of elements in dictionary B
\safemath{\elemA}{{n_a}}	%number of elements chosen from dictionary A
\safemath{\elemB}{{n_b}}	%number of elements chosen from dictionary B
\safemath{\resA}{\matR_a}	%restriction map to elements of dictionary A
\safemath{\resB}{\matR_b}	%restriction map to elements of dictionary B
\safemath{\subD}{\matS} %subdictionary
\safemath{\subA}{\matS_a} %subdictionary part of A
\safemath{\subB}{\matS_b} %subdictionary part of B
\safemath{\dicta}{\matA} 	% first subdictionary
\safemath{\dictb}{\matB} 	% second subdictionary
\safemath{\hollowS}{H}
\safemath{\hollowA}{H_a}
\safemath{\hollowB}{H_b}
\safemath{\cross}{Z}
\safemath{\coh}{\mu_d}			% coherence dictionary
\safemath{\coha}{\mu_a}			% coherence first subdictionary
\safemath{\cohb}{\mu_b}			% coherence second subdictionary
\safemath{\mubs}{\nu}	%block sub-coherence
\safemath{\cohm}{\mu_m} %mutual coherence
\safemath{\dictset}{\setD}	% set of dictionaries
\safemath{\dictsetp}{\dictset(\coh,\coha,\cohb)}	% set of dictionaries parametrized
\safemath{\dictsetgen}{\dictset_\text{gen}}
\safemath{\dictsetgenp}{\dictsetgen(\coh)}
\safemath{\dictsetonb}{\dictset_\text{onb}}
\safemath{\dictsetonbp}{\dictsetonb(\coh)}

\safemath{\leftside}{U}
\safemath{\rightsideA}{R_a}
\safemath{\rightsideB}{R_b}

\safemath{\indexS}{\setI_S} %set of indices participating in sub-dictionary S

\safemath{\na}{n_a}			% cardinality of set of linearly independent columns of first dictionary
\safemath{\nb}{n_b}			% cardinality of set of linearly independent columns of second dictionary
\safemath{\coeffa}{p_i}	%coefficients for columns of A
\safemath{\coeffb}{q_j}	%coefficients for columns of B
\safemath{\seta}{\setP}		% set of linearly independent columns of A
\safemath{\setb}{\setQ}     % set of linearly independent columns of B
\safemath{\setw}{\setW}	%set of n largest elements of w
\safemath{\setz}{\setZ}	%set of L-n largest elements of z
\safemath{\cola}{\veca}		% generic element of the dictionary A
\safemath{\colb}{\vecb}		% generic element of the dictionary B
\safemath{\cold}{\vecd}		% generic element of the dictionary D
\safemath{\inputvec}{\vecx} 	%coefficient vector (input)
\safemath{\error}{\vece}	%error vector
\safemath{\noiseout}{\vecz} 	%noisy output vector
\safemath{\inputvecel}{x}
\safemath{\inputveca}{\vecx_a}
\safemath{\inputvecb}{\vecx_b}
\safemath{\outputvec}{\vecy}	%output of Dictionary
\safemath{\lambdamin}{\lambda_{\mathrm{min}}}
\safemath{\elltwo}{\ell_2}
\safemath{\ellone}{\ell_1}
\safemath{\ellzero}{\ell_0}
\safemath{\ellinf}{\ell_\infty}
\safemath{\ellinftilde}{\ell_{\widetilde\infty}}
\safemath{\licard}{Z(\coh,\coha,\cohb)}
\safemath{\xsol}{\hat{x}}
\safemath{\xbord}{x_b}		%Solution at the border
\safemath{\xstat}{x_s}		%Solution stationary in l0 prob
\safemath{\xstatLone}{\tilde{x}_s}
\safemath{\order}{\mathcal{O}} %order notation (big O)
\safemath{\scales}{\Theta} %scales as
\safemath{\ones}{\mathbf{1}} %all ones matrix
\safemath{\zeroes}{\mathbf{0}} %all zeroes matrix
\safemath{\thlone}{\kappa(\coh,\cohb)} %treshold l1 problem
\safemath{\constoneA}{\delta} %constant in l1 theorem to save space
\safemath{\constoneB}{\epsilon} %constant in l1 theorem to save space
\safemath{\nlarge}{L}				   %num large elements
\safemath{\sumlarge}{S_\nlarge}
\safemath{\maxlarger}{P_\nlarge}	   % maximum in Gribonval and Nielsen
\safemath{\Pzero}{\textrm{P0}}	
\safemath{\Pone}{\textrm{P1}}
\safemath{\vecfir}{\vecw}			 % \vecv element of the kernel of the dictionary, \vecv=[\vecfir \vecsec]
\safemath{\vecsec}{\vecz}
\safemath{\elvecfir}{w}              % element of vecfir
\safemath{\elvecsec}{z}				 % element of vecsec
\safemath{\nlargefir}{n}
\safemath{\normout}{\gamma}
\safemath{\auxfun}{h}
\safemath{\supp}{\textrm{supp}}%support

\safemath{\indexa}{\ell}
\safemath{\indexb}{r}
\safemath{\indexc}{i}
\safemath{\indexd}{j}

\safemath{\project}{P}%projector

\safemath{\firstslotset}{\setU_1}  % set of UEs for first slot
\safemath{\secondslotset}{\setU_2} % set of UEs for second slot
\safemath{\randomset}{\setS} % generic random scheduling

\newcommand*\squeezespaces[1]{% %% <- #1 is a number between 0 and 1
  \thickmuskip=\scalemuskip{\thickmuskip}{#1}%
  \medmuskip=\scalemuskip{\medmuskip}{#1}%
  \thinmuskip=\scalemuskip{\thinmuskip}{#1}%
  \nulldelimiterspace=#1\nulldelimiterspace
  \scriptspace=#1\scriptspace
}
\newcommand*\scalemuskip[2]{%
  \muexpr #1*\numexpr\dimexpr#2pt\relax\relax/65536\relax
} %% <- based on  https://tex.stackexchange.com/a/198966/156366

\usepackage{pifont}% http://ctan.org/pkg/pifont
\newcommand{\cmark}{\ding{51}}%
\newcommand{\xmark}{\ding{55}}%

\newcommand*{\fancyreflstlabelprefix}{lst}
\fancyrefaddcaptions{english}{%
  \providecommand*{\freflstname}{Listing}%
}
\frefformat{vario}{\fancyreflstlabelprefix}{%
  \freflstname\fancyrefdefaultspacing#1#3%
}

\begin{document}
\bstctlcite{IEEEexample:BSTcontrol}

\title{PyJama: Differentiable Jamming and\\Anti-Jamming with NVIDIA Sionna} 
\author{%
    \IEEEauthorblockN{Fabian Ulbricht, Gian Marti, Reinhard Wiesmayr, and Christoph Studer}\\
    \textit{Department of Information Technology and Electrical Engineering, ETH Zurich, Switzerland}\\
    \textit{email: fabianu@student.ethz.ch, marti@iis.ee.ethz.ch, wiesmayr@iis.ee.ethz.ch, studer@ethz.ch}\\
    \thanks{The work of GM, RW, and CS was supported in part by an ETH Zurich Research Grant.}
    \thanks{The authors thank J. Hoydis, S. Cammerer, and F. A. Aoudia\newline
    for comments and discussions on NVIDIA Sionna. The authors \vspace{-8.8mm}\newline
    also thank C. Dick for helpful comments and discussions.\hfill \includegraphics[width=12mm]{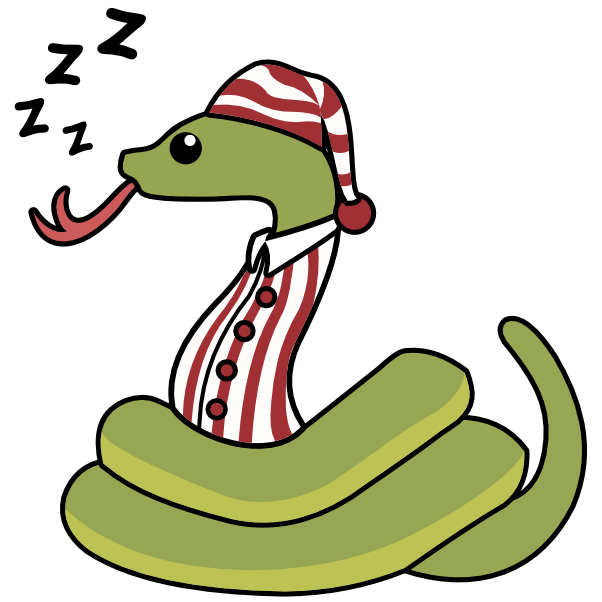}}
}

\maketitle

\begin{abstract}
Despite extensive research on jamming attacks on wireless communication systems, the potential of machine learning for amplifying the threat of such attacks, or our ability to mitigate them, remains largely untapped. A key obstacle to such research has been the absence of a suitable framework. To resolve this obstacle, we release PyJama, a fully-differentiable open-source library that adds jamming and anti-jamming functionality to NVIDIA~Sionna. 
We demonstrate the utility of PyJama (i) for realistic MIMO simulations by showing examples that involve forward error correction, OFDM waveforms in time and frequency, realistic channel models, and mobility; and (ii) for learning to jam. Specifically, we use stochastic gradient descent to optimize jamming power allocation over an OFDM resource grid. 
The learned strategies are non-trivial, intelligible, and effective. 

\end{abstract}

\section{Introduction}

Jamming attacks on wireless communication systems are a growing concern and have been studied extensively \cite{pirayesh2022jamming}. Since machine learning (ML) had a transformative impact on communication engineering \cite{OShea2017a, erpek2020deep}, it might be expected to enhance our understanding of jamming attacks and facilitate new ways of mitigating such attacks. However, the potential of ML for either jamming or anti-jamming remains largely untapped, mainly due to the lack of a suitable framework.

\subsection{Contributions}
In order to facilitate the use of ML for research in jamming and anti-jamming, we release PyJama, a fully-differentiable open-source library that adds jamming and anti-jamming (with a focus on MIMO anti-jamming) functionality to \mbox{NVIDIA Sionna \cite{Hoydis2022}.}
The combination of Sionna and PyJama enables realistic link-level simulations of wireless communication systems under jamming attacks, including aspects such as jammer-mitigating or jammer-oblivious MIMO receivers; forward error correction (FEC); orthogonal-frequency division multiplexing (OFDM) waveforms in the time or frequency domains; geometric channel models \cite{Hoydis2023}; transmitter, receiver, and jammer mobility; and much more. In \fref{sec:api_overview_examples}, we provide a brief overview of PyJama’s application programming interface (API) and showcase some of its capabilities via simulation examples. 
Extensive documentation and Jupyter notebooks for all of our experiments are available on \href{http://pyjama.ethz.ch}{\texttt{pyjama.ethz.ch}}.
We then leverage PyJama for learning to jam (against traditional receivers as well as against basic anti-jamming MIMO receivers), 
by using stochastic gradient descent to optimize jamming power allocation over an OFDM resource grid. 
We also propose suitable jammer loss functions for uncoded and coded communication systems. 
The learned jamming strategies are nontrivial, intelligible, and highly~effective. 

\subsection{Prior Work}
Prior studies on jamming and anti-jamming strategies have often used game-theoretic approaches, 
where transmitters and jammers correspond to actors in Bayesian games~\cite{sagduyu2011jamming} or Stackelberg games~\cite{yang2013coping, shen2021beam} that model basic communication schemes.  
Other studies have used reinforcement learning (RL),  
such as \mbox{Q-learning} \cite{han2017two, liu2018anti, hoang2022multiple}, bandit optimization~\cite{kim2020reinforcement}, 
or adversarial ML \cite{erpek2018deep, erpek2020deep}.
All these works rely on basic models for the received signal-to-interference-plus-noise ratio (SINR) 
and are unable to account for the specifics of real-world communication systems such as waveforms,~channel~coding,~etc. 

Spatial filtering has recently gained traction as a powerful tool for jammer mitigation in MIMO systems\cite{pirayesh2022jamming},
but it is known that smart jammers can evade simple mitigation schemes at the receiver \cite{marti2023mitigating}. 
While \cite{shen2021beam, hoang2022multiple, kim2020reinforcement} consider jamming in MIMO systems, 
they do not leverage ML for learning adversarial jamming strategies against MIMO anti-jamming receivers, making our work the first to do so.

\subsection{Prerequisites} \label{sec:prereq}
PyJama is designed with a focus on MIMO (anti-)jamming. 
A jammed MIMO communication system can be modeled in discrete time as
\begin{equation}\textstyle
    \vecy[k] = \sum_{\ell=0}^{L-1} \big(\matH[k,\ell] \vecs[{k-\ell}] + \matJ[k,\ell] \vecw[{k-\ell}]\big) + \vecn[k], \label{eq:io}
\end{equation}
where $\bmy[k]\in\opC^B$ is the time-$k$ receive signal at a \mbox{$B$-antenna} receiver; 
$\bms[k]\in\opC^U$ and $\bmw[k]\in\opC^I$ are the time-$k$ transmit signals at one or multiple transmitters 
and one or multiple jammers with a total of $U$ and $I$ antennas, respectively; 
$\matH[k,\ell]\in\opC^{B\times U}$ and $\matJ[k,\ell]\in\opC^{B\times I}$ represent the impulse response matrices 
at time $k$ and delay~$\ell<L$ of the transmitter and jammer channels, respectively;  
and $\bmn[k]\in\opC^B$ is white Gaussian noise. 
In flat fading systems, the input-output relation in \eqref{eq:io} can be simplified to $\bmy[k]=\bH\bms[k]+\bJ\bmw[k]+\bmn[k]$.
A~receiver that uses a conventional data detector, such as an LMMSE equalizer
without any anti-jamming capability, will experience significantly increased error rates under jamming. 
In contrast, an anti-jamming receiver can mitigate the jamming interference through spatial filtering, for instance 
by projecting the receive signal onto the orthogonal complement of the jammer's subspace (``POS'') to null the jammer,\footnote{
In the flat-fading case, the corresponding projection matrix is \mbox{$\bI_B-\bJ\pinv{\bJ}$.}
}or by using an LMMSE equalizer that treats the jammer interference as spatially correlated noise (``IAN-LMMSE'')
\cite{marti2023universal}.

\section{PyJama: API and Examples}\label{sec:api_overview_examples}

We release PyJama, a TensorFlow-based open-source library that adds fully differentiable jamming and MIMO anti-jamming functionality 
to NVIDIA Sionna \cite{Hoydis2022}. 
The library, extensive API documentation, tutorials, and simulation code to reproduce the results of this paper are available on 
\href{http://pyjama.ethz.ch}{\texttt{pyjama.ethz.ch}}.

\subsection{API Overview}
\fref{fig:pyjama_pipeline} shows a basic PyJama-Sionna simulation pipeline of a coded OFDM system (forked arrows indicate multiple options). 

\subsubsection{Jammers}
PyJama enables the simulation of jammed communication systems in the time as well as frequency domain. 
Frequency-domain simulations are computationally more efficient while time-domain simulations can account for effects
that are otherwise ignored, such as MIMO-OFDM jammers that do not transmit a cyclic prefix (see~\fref{sec:time_domain_ofdm}). 
In both domains, jammers are implemented as communication blocks that superimpose the jamming interference
on the output of an unjammed wireless channel (see~\fref{fig:pyjama_pipeline}). 
This approach enables the use of potentially 
different channel models for the jammers and the transmitters (see~\fref{sec:mobility}). 
PyJama features jammers that transmit symbols drawn uniformly from a complex disk, complex Gaussian symbols, 
symbols from a QAM constellation, or symbols provided by a Python callable (see \fref{fig:resource_grid}). 
While time-domain jammers currently only support barrage (i.e., stationary) jamming, frequency-domain jammers
can be configured as barrage jammers, pilot jammers, data jammers, or sparse (in time or frequency) jammers (see \fref{fig:resource_grid}). 

\subsubsection{Receivers}
\begin{figure}[tp]
    \centering
    \includegraphics[width=.99\columnwidth]{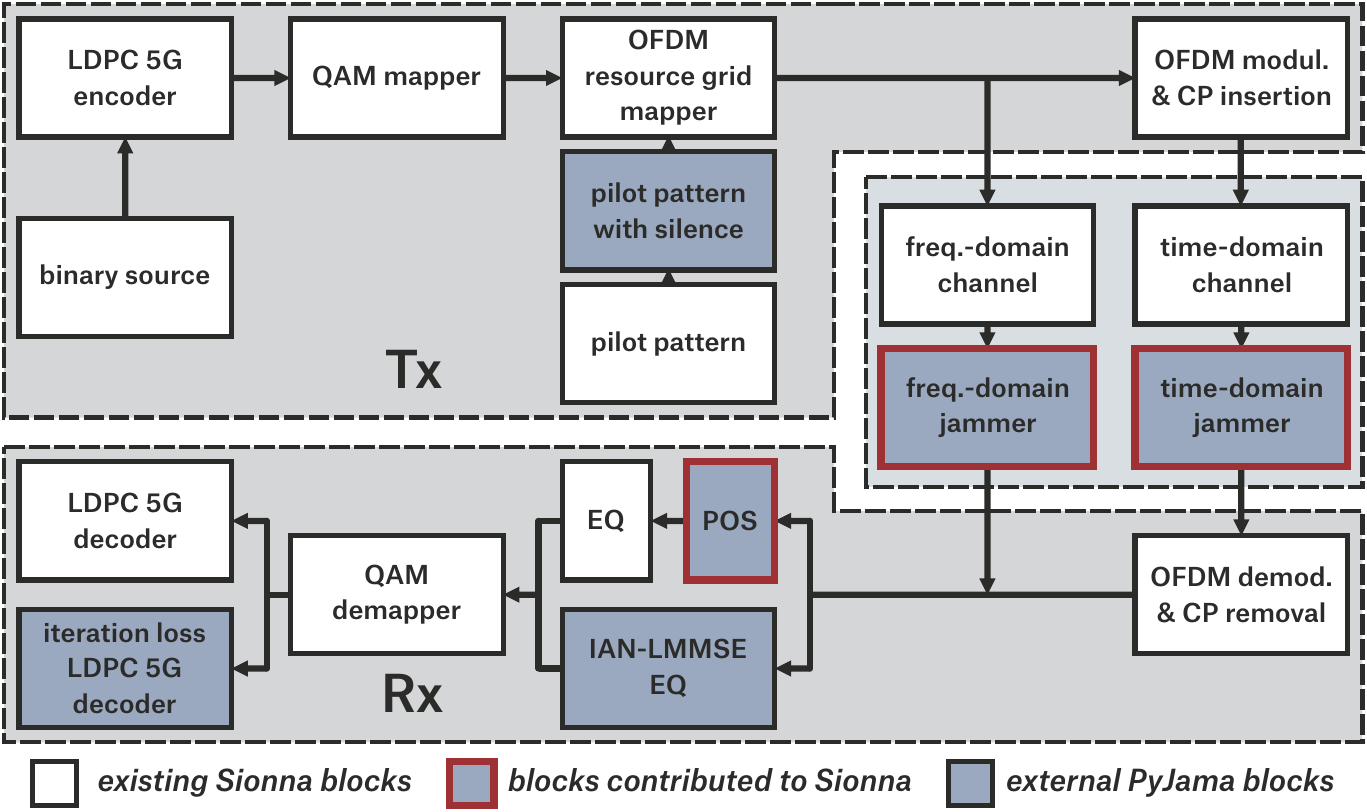}
    \vspace{-5mm}
    \caption{PyJama simulation pipeline (channel estimation not shown). In Sionna, the frequency-domain
    and time-domain jammers are implemented as (generalized) frequency-domain and time-domain interference 
    classes,~respectively.}
    \label{fig:pyjama_pipeline}
\end{figure}

PyJama implements the two anti-jamming receive methods POS and IAN-LMMSE (see \fref{sec:prereq}). 
While POS can be put in front of any Sionna equalizer, IAN-LMMSE unifies jammer mitigation 
and data detection, and so is a complete equalizer by itself (see \fref{fig:pyjama_pipeline}).
POS and IAN-LMMSE can either operate using perfect channel state information (CSI) for the transmitters and the jammers, 
or they can estimate the CSI (which requires new pilot patterns for jammer estimation; 
see below). For learning in coded systems---not necessarily restricted to (anti-)jamming applications---PyJama
also implements an iteration-loss LDPC decoder (see~\fref{fig:pyjama_pipeline}).

\subsubsection{Transmitters}
\begin{figure}[tp]
    \centering
    \includegraphics[width=\columnwidth]{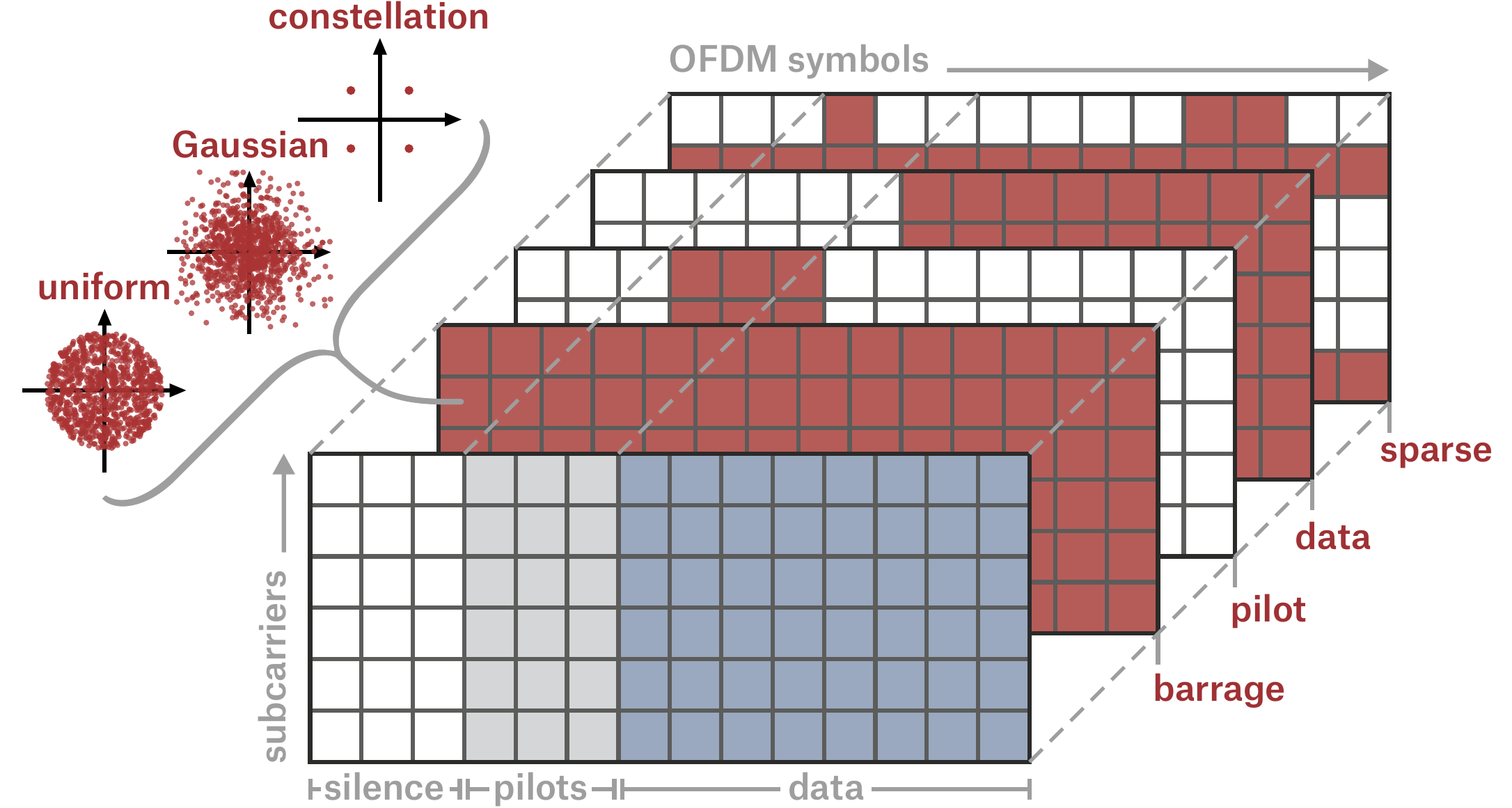}
    \vspace{-5mm}
    \caption{Foreground: Illustration of an OFDM resource grid that uses the \texttt{\footnotesize PilotPatternWithSilence}. Background: Corresponding illustration of the different types of frequency domain jammers (barrage jammer, pilot jammer, data jammer, sparse jammer). The jammer transmit symbols can be drawn from a uniform distribution, a Gaussian distribution, or a QAM constellation.}
    \label{fig:resource_grid}
\end{figure}

In real-world wireless systems, CSI has to be obtained through channel estimation. 
Since the anti-jamming receivers of PyJama depend on the jammers' CSI,
PyJama implements the \texttt{\small PilotPatternWithSilence}, which 
is a wrapper for a Sionna pilot pattern that adds ``silent'' resource elements (REs) 
to an OFDM resource grid (cf. \fref{fig:resource_grid}). 
During such silent REs, the transmitters
are idle and so enable the receiver to estimate the jammers' channel matrix up to complex-valued scale factors
(provided that the jammers jam these~REs).

\subsection{Examples}\label{sec:examples}

Since PyJama builds on Sionna, it enables convenient, fast, and realistic simulation of 
(potentially anti-jamming) communication systems under jamming attacks.
In the following examples, we showcase just a handful of possibilities. 
All of the included examples consider the uplink of a 5G NR-like multiuser (MU) MIMO system. 
Unless stated otherwise, the simulations use a 3GPP urban micro (UMi) channel model for both the user equpiments (UEs) 
as well as the jammers. The basestation (BS) has a dual-polarized uniform linear array (ULA) with a total of 16 antennas.
We consider four single-polarized single-antenna UEs (with perfect power control) and one single-polarized single-antenna jammer. 
We use OFDM at a 3.5\,GHz carrier frequency and 128 consecutive subcarriers spaced at 30\,kHz.
The resource grid contains 14 OFDM symbols. For each subcarrier, the first 4 of these symbols are silent (for jammer CSI estimation), 
the subsequent four symbols contain one-hot pilots (for UE CSI estimation), 
and the remaining six symbols contain QPSK data symbols. 
The jammer is a barrage jammer that transmits i.i.d.\ uniformly sampled symbols from the complex disk. 
In our experiments, if the receiver uses anti-jamming, then it does so by using POS (see \fref{sec:prereq}) in combination 
with a least-squares (LS) channel estimator and an LMMSE data detector; otherwise, the receiver uses conventional LS channel estimation and LMMSE data detection. 
In coded simulations, we use a 5G rate-1/2 LDPC code with a block length of 1536 bits, 
where the decoder uses max-product decoding with 20 flooding iterations.

\subsubsection{Bit Error Rate (BER) and Block Error Rate (BLER) Simulations} \label{sec:basic}
\begin{figure}
    \begin{minipage}[b]{.49\linewidth}
        \centering
        \includegraphics[width=4.5cm]{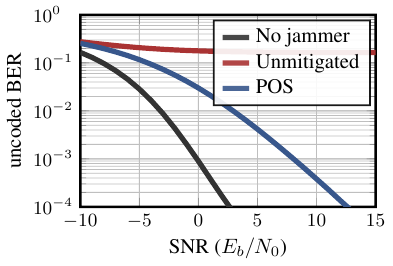}
        \vspace{-0.2cm}
    \end{minipage}
    \begin{minipage}[b]{.49\linewidth}
        \centering
        \includegraphics[width=4.5cm]{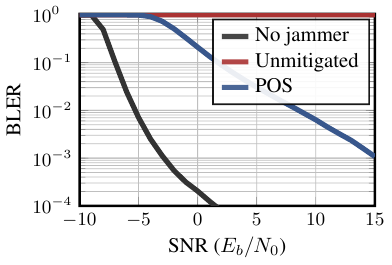}
        \vspace{-0.2cm}
    \end{minipage}
    \hfill
    \vspace{-6mm}
    \caption{Left: Uncoded bit error rates (BERs) vs. SNR for basic MIMO jamming and anti-jamming.
    Right: Corresponding block error rates (BLERs).}
    \label{fig:ber_vs_bler}
\end{figure}

Our first example consists of basic BER and BLER simulations (as a function of signal-to-noise ratio $E_b/N_0$)
that compare the performance between (i) a conventional receiver that is not attacked by jamming, 
(ii) a conventional receiver attacked by two 2-antenna jammers that jam at twice
the power of the average UE, 
and (iii) a POS anti-jamming receiver.
\fref{fig:ber_vs_bler} shows that the unjammed system has the best performance, while the jammed system with 
unmitigated receiver suffers from a high error floor that cannot be remedied by the LDPC code (in fact, the BLER is equal to one). 
The anti-jamming receiver is able to mitigate the jammer and resolve the error floor. 
The performance loss compared to the unjammed system is due to imperfectly estimated jammer CSI
and the loss of four degrees of freedom incurred by the POS projection. 

\subsubsection{Transmitter and Jammer Mobility} \label{sec:mobility}
\begin{figure}
    \begin{minipage}[b]{.49\linewidth}
        \centering
        \includegraphics[width=4.5cm]{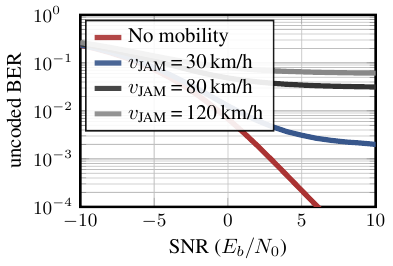}
        \vspace{-0.2cm}
    \end{minipage}
    \begin{minipage}[b]{.49\linewidth}
        \centering
        \includegraphics[width=4.5cm]{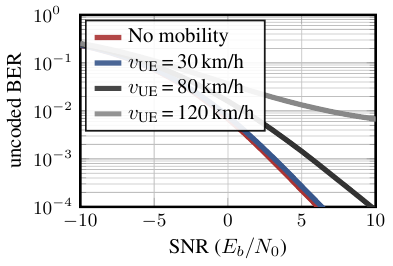}
        \vspace{-0.2cm}
    \end{minipage}
    \hfill
    \vspace{-6mm}
    \caption{Left: POS mitigation effectiveness for different levels of jammer mobility. 
    Right: POS mitigation effectiveness for different levels of UE mobility. }
    \label{fig:velocity_doppler}
\end{figure}

Studies on MIMO jammer mitigation usually operate, at least implicitly, with a block fading channel model, 
so that a CSI estimate from some point in time remains valid for a certain period. 
However, real-world wireless channels are not block fading. 
In \fref{fig:velocity_doppler}, we use PyJama to evaluate the impact of jammer and UE mobility, 
which models continually evolving communication channels, on MIMO anti-jamming against a
single-antenna jammer that jams at 20\,dB higher power than the average UE. 
The results show that, at low velocities, the impact of UE mobility is negligible 
while the impact of jammer mobility is not. 
At high velocities, UE mobility becomes a problem as well, but it is not as critical as jammer mobility.

\subsubsection{Time-Domain OFDM Jamming}\label{sec:time_domain_ofdm}
In our third example, we simulate a single-antenna OFDM-MIMO jammer (and its mitigation) in the time domain. 
It has recently been shown~\cite{Marti2023a} that a jammer that violates the OFDM protocol by not transmitting 
a cyclic prefix (CP) appears on each received subcarrier as high-rank interference---and not as rank-one interference, 
as an OFDM compliant jammer theoretically would. Specifically, \cite{Marti2023a} has shown that the rank of the receive interference
per subcarrier for a CP-violating jammer can be up to $\min\{B,L\}$, where~$B$ is the number of receive antennas, 
and~$L$ is the jammer's number of channel taps in the time domain. Consequently, such jammers
cannot be completely mitigated by nulling a single spatial dimension. So far, however, this effect 
has only been verified on simplistic channel models with i.i.d.\ Rayleigh fading channel taps, 
and its impact in more realistic models has not been investigated.

\begin{figure}
    \begin{minipage}[b]{.49\linewidth}
        \centering
        \includegraphics[width=4.5cm]{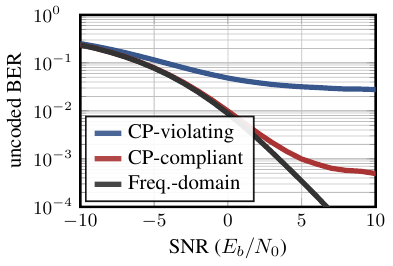}
        \vspace{-0.2cm}
    \end{minipage}
    \begin{minipage}[b]{.49\linewidth}
        \centering
        \includegraphics[width=4.5cm]{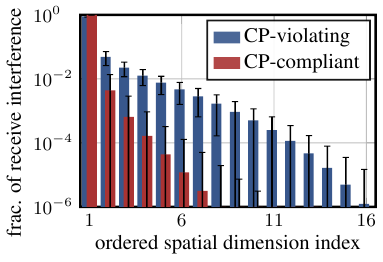}
        \vspace{-0.2cm}
    \end{minipage}
    \hfill
    \vspace{-6mm}
    \caption{Left: Error rates of CP-compliant and CP-violating single-antenna jammers under one-dimensional POS mitigation. 
    Right: Normalized histogram of the fraction of receive interference per spatial dimension (in descending order);
    the jammer simulated in the frequency domain is not shown.}
    \label{fig:ofdm_vs_time_domain}
\end{figure}

In \fref{fig:ofdm_vs_time_domain}, we simulate such a scenario. 
We consider a channel with 20\,MHz bandwidth and 30\,kHz subcarrier spacing (=\,667 subcarriers);  
the CP-length is 2.35\,$\mu$s (=\,47~channel taps). 
The left plot shows the error rate for a POS receiver that nulls the strongest spatial interference 
dimension (based on perfect~CSI) for (i) a CP-violating single-antenna jammer simulated in the time domain, 
(ii) a CP-compliant single-antenna jammer simulated in the time domain, 
and (iii) a single-antenna jammer simulated in the frequency domain. 
All jammers jam at 20\,dB higher power than the average UE. 
The results show that, for the first jammer, the receive interference is not restricted to a one-dimensional subspace,
which leads to poor performance even after nulling the strongest interference dimension. 
The frequency-domain simulation ignores effects related to imperfect CPs and so the single-antenna jammer is perfectly
nulled by the POS receiver, since it only causes rank-one interference per subcarrier. 
In theory, if the CP is not shorter than the channel's impulse response, the CP-compliant jammer should
behave identically to the frequency-domain simulation. However, the error floor induced by the CP-compliant 
jammer indicates that, at least in some cases, the delay spread exceeds the CP length.
The right plot in \fref{fig:ofdm_vs_time_domain} supports these conclusions by
showing normalized histograms of the ordered singular values of the receive jammer interference, 
which correspond to the relative fraction of receive interference per spatial dimension: 
A CP-compliant single-antenna jammer would theoretically cause rank-one interference and thus have 
only one singular value with non-zero distribution. But we see that the CP-violating jammer has 16 
singular values with non-zero distribution, indicating full-rank receive interference (since the BS has only 16 antennas). 
Even the CP-compliant jammer has multiple singular values with non-zero distribution since the channel impulse response length 
sometimes exceeds the CP-length. The jammer simulated in the frequency domain is only rank one, and that rank's  
fraction of receive interference is equal to one (not shown in \fref{fig:ofdm_vs_time_domain}).

\section{Learning to Jam}\label{sec:trainable_jammers}

While PyJama is equally suited for learning to mitigate jammers, we now use it for \emph{learning to jam,} with the goal of improving our understanding of effective jamming attacks. Specifically, our goal is to learn jammers
that maximize the error rate for a given power budget $\rho_{\max}$ (compared to the average UE), against
traditional MIMO receivers as well as POS-based anti-jamming. 
To this end, we use empirical risk minimization (ERM) based on stochastic gradient descent.

\subsection{Fundamentals of Jammer Learning}

In order to learn how to jam, one needs a parametrized class of jamming strategies 
and a model that is differentiable with respect to the parameters of that class. In our case, the jamming strategies are parametrized by a vector $\bm{\rho}$ that parametrizes
the jamming power allocation (see \fref{sec:ml_model});  
and the model is a PyJama pipeline (as in \fref{fig:pyjama_pipeline}) simulating a jammed communication system, 
combined with a loss function~$\ell$ that captures the empirical error. 
Specifically, PyJama simulates the transmission of information bits $\bmb\in\{0,1\}^N$ 
with corresponding soft-output estimates $\hat\bmb\in[0,1]^N$ at the receiver.\footnote{
The soft-outputs $\hat\bmb=(\hat{b}_1, \dots, \hat{b}_N)\in[0,1]^N$ represent bit estimates in the probability domain, not LLRs. 
Specifically, $\hat{b}_n$ represents the estimate of the probability $P(b_n=1)$ that the $n$th transmitted bit 
is one. 
}
The function $\ell(\bmb,\hat\bmb)$ measures the error between the true bits $\bmb$ and their estimates~$\hat\bmb$.
Due to PyJama's diferentiability, $\ell(\bmb,\hat\bmb)$ is differentiable with respect to $\bm{\rho}$ 
as long as it is differentiable with respect to $\hat\bmb$.
By defining the jammer power allocation vector $\bm{\rho}$ as a trainable TensorFlow variable,
one can therefore use TensorFlow's autodiff capabilities to learn how to jam.  

A commonly used loss function for ML in communications is the binary cross entropy (BCE). 
While the BCE is well-suited for (bit) error rate \emph{minimization} \cite{Wiesmayr2023}, 
it is unsuited for error rate \emph{maximization} (which is our goal when learning to jam):
The BCE loss $\ell_{\text{BCE}}(\bmb,\hat\bmb)$ is unbounded on
$\{0,1\}^N\times[0,1]^N$ and---in particular---can be arbitrarily large even when only one bit estimate is incorrect, 
meaning that BCE fails to capture the \emph{rate} of errors. 
This defect is avoided by loss functions such as the L1 loss and the mean squared error (MSE); see \fref{tbl:loss_fun}.
Both of these have proven to work well for the task at hand.

\subsection{Training Model} \label{sec:ml_model}
We simulate a MU-MIMO OFDM system as in \fref{sec:examples} in the frequency domain, with the following differences: 
(i)~depending on the experiment, the number of UEs (and~the corresponding length of the pilots) varies between one and eight;
(ii) depending on the experiment, the receiver is either the conventional MIMO receiver 
(in which case the resource grid contains no silent symbols; see \fref{fig:learned_resource_grid}) 
or the POS receiver from \fref{sec:basic}; 
(iii) the jammer transmits independent Gaussian symbols, where the symbols transmitted during the~$i$th OFDM symbol
have trainable power $\rho_i$ subject to a power constraint $\|\bm{\rho}\|_1/N_s\triangleq\frac{1}{N_s}\sum_{i=1}^{N_s} \rho_i \leq \rho_{\max}$, 
where $N_s=14$ is the number of OFDM symbols per resource grid; see \fref{fig:learned_resource_grid}.
When considering uncoded systems, we use  the L1 loss from \fref{tbl:loss_fun}. 
When considering coded systems (with an LDPC code as described in \fref{sec:examples}), 
we use an iteration loss LDPC 5G decoder based on the L1 loss of the information bits (cf. \fref{fig:pyjama_pipeline}), 
which returns soft estimates of the information bits after every iteration, and 
which we use as a multi-loss function \cite{Nachmani2018}.

\bgroup
\setlength{\tabcolsep}{0.5em}
\def\arraystretch{1.5}
\begin{table}[tp]
\begin{center}
\caption{Unsuitable (\xmark) and Suitable (\cmark) Jammer Loss Functions \vspace{-.5cm}}
\label{tbl:loss_fun}
\vspace{3mm}
\begin{tabular}{ c c c}
\hline
 \bf\qquad BCE \xmark & \bf\qquad L1 \cmark & \bf\qquad MSE \cmark \\[.8ex]
 $ \tiny \squeezespaces{0.25} \displaystyle \!\!\!\!-\!\!\sum_{n=1}^N \frac{b_n\log\hat b_n + (1-b_n)\log(1-\hat b_n)}{N}$ 
 & $ \tiny \squeezespaces{0.25} \displaystyle \!\!\sum_{n=1}^N \frac{|b_n - \hat b_n|}{N}$
 & $ \tiny \squeezespaces{0.25} \displaystyle \!\!\sum_{n=1}^N \frac{(b_n - \hat b_n)^2}{N}\!\!\!\!$ \vspace{1mm} \\ 
 \hline
\end{tabular}
\end{center}
\vspace{-3mm}
\end{table}

\begin{figure}[tp]
    \centering
    \includegraphics[width=\columnwidth]{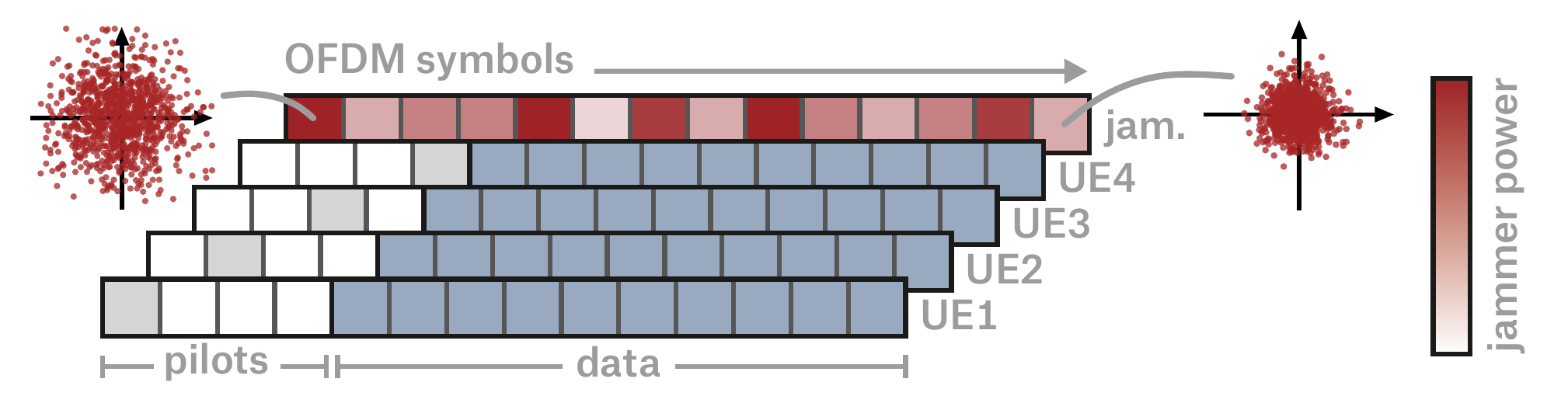}
    \vspace{-6mm}
    \caption{Illustration of the UE resource grid for a classical MIMO receiver (only one subcarrier shown) 
    with 4 UEs and one-hot pilots, and of a trainable~jammer.}
    \label{fig:learned_resource_grid}
\end{figure}

\begin{figure*}[tp]
    \centering
    \hspace{-2.5mm}
    \includegraphics[width=1.002\textwidth]{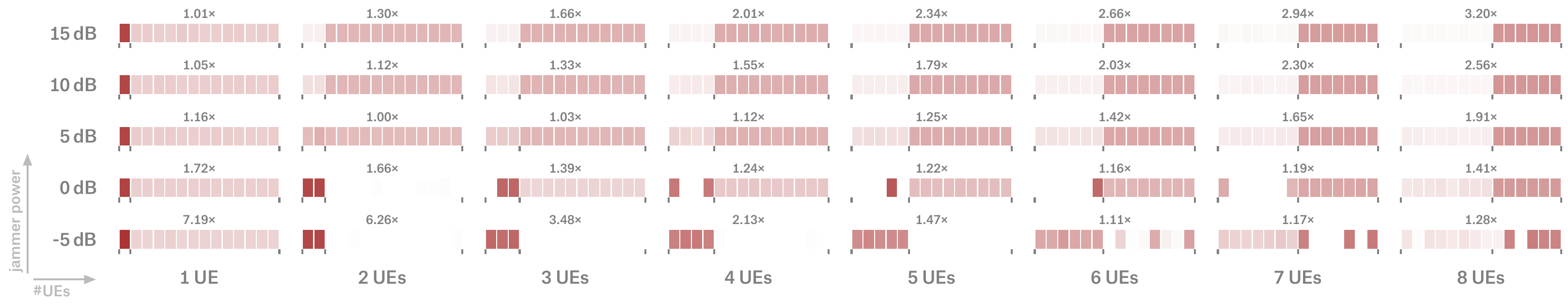}
    \vspace{-7mm}
    \caption{Learned jammer power allocation for different jammer powers and numbers of UEs. 
    The displayed factors indicate the BER increase at an SNR of 5\,dB between the learned jammer and a uniform barrage jammer.
    The ticks indicate the edges of the pilots (whose lengths equal the number of UEs) and data phases.
    }
    \vspace{-3mm}
    \label{fig:trained_resource_grid}
\end{figure*}

For optimization, we use Adam with a learning rate of $10^{-3}$.
We train for 5000 gradient steps, each of which processes a batch of
64 transmitted OFDM frames at an SNR of 0\,dB.

\subsection{Learning Results}

\subsubsection{Conventional Receiver, Uncoded Communications}
\fref{fig:trained_resource_grid} shows the learned jammer power 
allocation for different jammer power budgets and different numbers of UEs. 
The results show that the learned strategy depends heavily on the specific setting.

In single-UE settings, the jammer focuses its energy primarily on the pilot symbol. 
In few-UE settings, the learned strategies differ depending on the power budget:
a weak jammer jams primarily the pilots, but often targets one or two UEs
(since one-hot pilots are used, jamming, e.g., the third OFDM symbol affects 
the channel estimate, and thus the error rate, of the third UE);
a strong jammer, in contrast, learns to focus the energy 
on the data symbols, which affects all UEs equally. 
Finally, in many-UE settings, the jammer focuses its energy primarily on the 
data symbols, irrespective of its power budget. 

In terms of effectiveness, we note that the learned jammer always
outperforms a uniform barrage jammer: 
at an SNR of 5\,dB, the BER for the trained jammer exceeds
the BER of an equally strong barrage jammer by factors between 1.00 and 7.19. 
The effectiveness of jammer learning can also be quantified in terms of a 
power gain: The left plot in \fref{fig:trained_jammers_gains_ber_bler} shows
that a learned jammer with a power budget of $\rho_{\max}=-$5\,dB 
is as effective as a barrage jammer with a power of $-$2.5\,dB, 
indicating an ``effectiveness gain'' of 2.5\,dB. 
Moreover, a learned jammer with $\rho_{\max}=\,$10\,dB is more
effective than a barrage jammer with a power of 30\,dB, indicating 
an improvement of more than 20\,dB.

\subsubsection{Conventional Receiver, Coded Communications}
The right plot in \fref{fig:trained_jammers_gains_ber_bler} shows that learning 
also improves the effectiveness of jamming against coded systems: 
while a barrage jammer with 0\,dB power only causes a BLER of around 7\% 
(at an SNR of 5\,dB), a learned jammer with 0\,dB power constraint 
can drive the BLER up to 90\% at the same SNR and, therefore, drive the throughput to almost zero. 

\begin{figure}
    \begin{minipage}[b]{.49\linewidth}
        \centering
        \includegraphics[width=4.5cm]{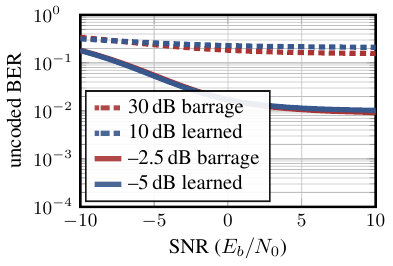}
        \vspace{-0.2cm}
    \end{minipage}
    \begin{minipage}[b]{.49\linewidth}
        \centering
        \includegraphics[width=4.5cm]{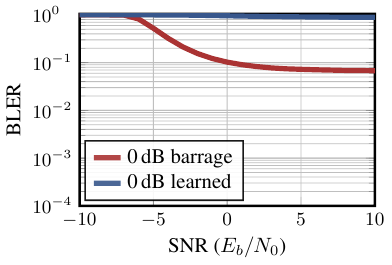}
        \vspace{-0.2cm}
    \end{minipage}
    \vspace{-6mm}
    \hfill
    \caption{Performance gains from jammer training against a conventional MIMO receiver with 4 UEs.
    Left: Jammer powers at which uniform barrage jammers are equally effective (in terms of BER) as learned jammers.
    Right: BLER against a 0\,dB uniform barrage jammer and a 0\,dB strong learned jammer.}
    \label{fig:trained_jammers_gains_ber_bler}
\end{figure}

\begin{figure}
    \begin{minipage}[b]{.49\linewidth}
        \centering
        \includegraphics[width=4.5cm]{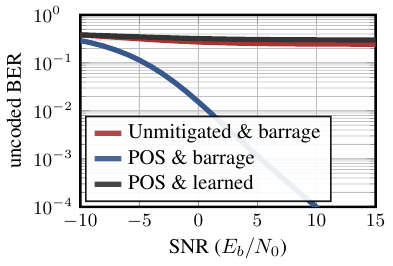}
        \vspace{-0.2cm}
    \end{minipage}
    \hfill    
    \begin{minipage}[b]{.49\linewidth}
        \centering
        \includegraphics[width=4.5cm]{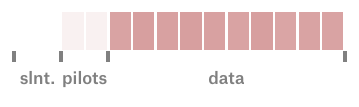}
        \vspace{0.7cm}
    \end{minipage}
    \vspace{-8mm}
    \caption{Performance gains from jammer training against a POS anti-jamming receiver with 4 UEs (left)
    and learned power allocation that does not jam during the silent symbols (right).}
    \label{fig:learned_vs_POS}
\end{figure}

\subsubsection{POS Anti-Jamming Receiver, Uncoded Communications}
Finally, \fref{fig:learned_vs_POS} show that jammers can learn how to bypass simple mitigation schemes. 
Here, we consider a POS anti-jamming receiver that estimates the jammer CSI from two silent symbols 
at the start of the resource grid. The right side of \fref{fig:learned_vs_POS} shows that the jammer 
learns to simply stay silent during those samples. Then, the POS receiver's estimate of 
the jammer CSI consist only of thermal noise, making anti-jamming completely ineffective 
(see the left side of \fref{fig:learned_vs_POS}).

\section{Conclusions and Future Work}
To unleash the potential of ML for understanding and mitigating jamming attacks, we have presented PyJama, 
a TensorFlow-based library that adds jamming and (MIMO) anti-jamming capability to NVIDIA Sionna.
Together with Sionna, PyJama enables realistic and fully-differentiable \mbox{link-level} simulations of 
wireless communication systems under jamming~attacks. 

We have leveraged PyJama to learn optimal jamming strategies in uncoded and coded MIMO-OFDM systems,
both against conventional receivers as well as against anti-jamming receivers. 
The learned jamming strategies are intelligible and highly effective. In particular, the jammers 
can learn to bypass simple MIMO mitigation schemes---this highlights the need for more sophisticated mitigation strategies.

% Generated by IEEEtran.bst, version: 1.14 (2015/08/26)

\end{document}